\documentstyle[12lomcon,cite,epsf]{article}
\begin{document}

\begin{flushright}
SHEP-06-03\\
\end{flushright}

\title{Spectrum of Higgs particles in the Exceptional
Supersymmetric Standard Model}

\author{ S.~F.~King\,\footnote{E-mail: \texttt{sfk@hep.phys.soton.ac.uk}.},\quad
S.~Moretti\,\footnote{E-mail: \texttt{stefano@hep.phys.soton.ac.uk}.},\quad
R.~Nevzorov\,\footnote{E-mail: \texttt{nevzorov@phys.soton.ac.uk}.}
\footnote{On leave of absence from the Theory Department, ITEP, Moscow, Russia.}
\footnote{Based on a talk presented by R.Nevzorov at the 12th Lomonosov Conference on 
Elementary Particle Physics, Moscow, Russia, 25-31 August 2005}
}

\address{School of Physics and Astronomy, University of Southampton,\\
Southampton, SO17 1BJ, U.K.}

\maketitle\abstracts{ 
We discuss the spectrum of Higgs bosons in the framework of the exceptional 
supersymmetric standard model. The presence of a $Z'$ and exotic 
particles predicted by the exceptional SUSY model allows the lightest Higgs 
particle to be significantly heavier than in the MSSM and NMSSM. When the mass 
of the lightest Higgs boson is larger than $135-140\,\mbox{GeV}$ the heaviest
scalar, pseudoscalar and charged Higgs states lie beyond the $\mbox{TeV}$
range.}

One of the strongest theoretical arguments in favour of softly
broken supersymmetry (SUSY) is associated with the partial unification of 
the standard model (SM) gauge interactions with gravity. However, the minimal supersymmetric 
standard model (MSSM) suffers 
from the $\mu$ problem. Indeed the superpotential of the MSSM contains one 
bilinear term $\mu \hat{H}_d\hat{H}_u$, so that $\mu$ 
is expected to be either zero or of the order of the Planck scale. 
At the same time, in order to get the correct pattern of electroweak symmetry 
breaking (EWSB), $\mu$ is required to be in the EW or TeV range. In the framework of the 
simplest extension of the MSSM --- the Next--to--Minimal Supersymmetric Standard 
Model (NMSSM) --- an `effective' $\mu$--term is generated dynamically. (For a review 
of the MSSM and NMSSM see e.g. \cite{1}.). 

A similar solution to the $\mu$ problem arises within superstring inspired
models based on the $E_6$ gauge group or its rank--6 subgroup 
$SU(3)_C\times SU(2)_W\times U(1)_Y\times U(1)_{\psi}\times U(1)_{\chi}$.
Two anomaly-free $U(1)_{\psi}$ and $U(1)_{\chi}$ symmetries of the
rank-6 model are defined by \cite{2}: $E_6\to SO(10)\times U(1)_{\psi},~
SO(10)\to SU(5)\times U(1)_{\chi}$. Near the string scale the rank-6 model 
can be reduced to an effective rank--5 model with only one extra gauge 
symmetry $U(1)'$:
\begin{equation}
U(1)'=U(1)_{\chi}\cos\theta+U(1)_{\psi}\sin\theta\,.
\label{2}
\end{equation}
If $\theta\ne 0\,\mbox{or}\,\,\pi$ the extra $U(1)'$ gauge symmetry 
forbids an elementary $\mu$ term but allows interaction $\lambda S H_d H_u$
in the superpotential. After EWSB the scalar component of the SM singlet 
superfield $S$ acquires a non-zero vacuum expectation value
(VEV) breaking $U(1)'$ and an effective 
$\mu$ term of the required size is automatically generated. 

In this article we explore the Higgs sector in the framework of a particular 
$E_6$ inspired supersymmetric model with an extra $U(1)_{N}$ gauge symmetry 
in which right handed neutrinos do not participate in the gauge interactions. 
($\theta=\arctan\sqrt{15}$). 
Only in this exceptional supersymmetric standard model (E$_6$SSM) 
\footnote{Similar abbreviation -- ESSM is used for the Extended Supersymmetric Standard
Model which involves two extra vector-like families [$16+\overline{16}$ of SO(10)] 
of quarks and leptons with masses of order of one TeV \cite{2a}. In contrast our
model involves three families of extra $10$'s of SO(10) which fill out three 
families of complete $27$'s of $E_6$ near the TeV scale (apart from right-handed 
neutrinos which are expected to have intermediate scale masses).
In our previous publications \cite{3}--\cite{4} we also used the acronym ESSM 
since we were unaware of the earlier work \cite{2a}.} right--handed 
neutrinos may be superheavy, sheding light on the origin of the mass hierarchy 
in the lepton sector \cite{3}--\cite{4}. The particle content of 
the E$_6$SSM involves three complete fundamental $27$ representations of $E_6$.
It ensures anomaly cancellation within each generation. In addition to the 
complete $27_i$ representations doublet and anti-doublet from extra $27'$ 
and $\overline{27'}$ can and must survive to low energies to preserve gauge 
coupling unification. Thus in addition to a $Z'$ the E$_6$SSM involves extra matter 
 that forms three $5+5^{*}$ representations of $SU(5)$ plus three 
$SU(5)$ singlets which carry $U(1)_N$ charges.

The E$_6$SSM Higgs sector includes two Higgs doublets $H_u$ and $H_d$ as well as a SM--like 
singlet field $S$ \cite{3}--\cite{4}. The Higgs effective potential can be written as 
\begin{equation} 
V=V_F+V_D+V_{soft}+\Delta V\,,\qquad\qquad\qquad\qquad\qquad\qquad\qquad
\label{40}
\end{equation}
\begin{equation}
\begin{array}{rcl}
V_F&=&\lambda^2|S|^2(|H_d|^2+|H_u|^2)+\lambda^2|(H_d H_u)|^2\,,\\[1mm]
V_D&=&\frac{g_2^2}{8}\left(H_d^\dagger \sigma_a H_d+H_u^\dagger \sigma_a
H_u\right)^2+\frac{{g'}^2}{8}\left(|H_d|^2-|H_u|^2\right)^2+\\[1mm]
&&+\frac{g^{'2}_1}{2}\left(\tilde{Q}_1|H_d|^2+\tilde{Q}_2|H_u|^2+\tilde{Q}_S|S|^2\right)^2\,,\\[1mm]
V_{soft}&=&m_{S}^2|S|^2+m_1^2|H_d|^2+m_2^2|H_u|^2+\biggl[\lambda
A_{\lambda}S(H_u H_d)+h.c.\biggr]\,, 
\end{array}
\label{3}
\end{equation} 
where $g'=\sqrt{3/5} g_1$; $g_2$, $g_1$ and $g'_1$ are $SU(2)_W$, $U(1)_Y$ and 
$U(1)_N$ gauge couplings; while $\tilde{Q}_1$, $\tilde{Q}_2$ and $\tilde{Q}_S$ 
are effective $U(1)_{N}$ charges of $H_d$, $H_u$ and $S$ respectively \cite{3}--\cite{4}. 
The couplings $g_2$ and $g'$ are known precisely. Assuming gauge coupling unification 
one can find that $g'_1(Q)\simeq g_1(Q)$ for any $Q< M_{\rm GUT}$ \cite{4}.

At tree--level the Higgs potential is described by the sum of the first three 
terms in Eq.~(\ref{40}). The last term $\Delta V$ represents the contribution
of loop corrections to the Higgs effective potential. In Eq.~(\ref{40}) 
$V_F$ and $V_D$ are the $F$ and $D$ terms. The soft SUSY breaking terms 
are collected in $V_{\rm soft}$. A simple counting shows that the E$_6$SSM Higgs sector 
contains only one additional singlet field and one extra parameter compared to 
the MSSM. Therefore it can be regarded as the simplest extension of the Higgs 
sector of the MSSM.

At the physical vacuum the E$_6$SSM Higgs fields develop the VEVs 
$\langle H_d\rangle =\frac{v_d}{\sqrt{2}},\,\langle H_u\rangle
=\frac{v_u}{\sqrt{2}}$ and $\langle S\rangle=\frac{s}{\sqrt{2}}$, 
thus breaking the $SU(2)_W\times U(1)_Y\times U(1)_N$ symmetry to $U(1)_{\rm EM}$, 
associated with electromagnetism. Instead of $v_d$ and $v_u$ it is more 
convenient to use $\tan\beta=\frac{v_u}{v_d}$ and $v=\sqrt{v_d^2+v_u^2}$, where
$v=246\,{\rm GeV}$. After EWSB two CP-odd and two charged Goldstone modes in the 
Higgs sector are absorbed by the $Z$, $Z'$ and $W^{\pm}$ gauge bosons so that only 
six physical degrees of freedom are left. They form one CP-odd and two charged Higgs 
states (as in the MSSM) with masses
\begin{equation}
m_A^2\simeq \frac{2\lambda^2 s^2 x}{\sin^2 2\beta}+O(M_Z^2)\,,\qquad 
m^2_{H^{\pm}}\simeq m_A^2+O(M_Z^2)
\label{4}
\end{equation}
respectively and three CP-even states (as in the NMSSM) with masses
\begin{equation}
\begin{array}{c}
m^2_{h_3}\simeq m_A^2+O(M_Z^2)\,,\qquad m^2_{h_2}\simeq g^{'2}_1\tilde{Q}_S^2s^2+O(M_Z^2)\,,
\end{array}
\label{5}
\end{equation}
\begin{equation}
\begin{array}{c}
m^2_{h_1}\simeq \frac{\lambda^2}{2}v^2\sin^22\beta+\frac{\bar{g}^2}{4}v^2\cos^22\beta+
g^{'2}_1 v^2\biggl(\tilde{Q}_1\cos^2\beta+\tilde{Q}_2\sin^2\beta\biggr)^2-\\[1mm]
-\frac{\lambda^4 v^2}{g^{'2}_1Q_S^2}\biggl(1-x+\frac{g^{'2}_1}{\lambda^2}
\biggl(\tilde{Q}_1\cos^2\beta+Q_2\sin^2\beta\biggr)Q_S\biggr)^2+...\,,
\end{array}
\label{6}
\end{equation}
where the auxiliary variable is $x=\frac{A_{\lambda}}{\sqrt{2}\lambda s}\sin 2\beta$.

At least one CP--even Higgs boson is always heavy preventing the distinction between 
the E$_6$SSM and MSSM Higgs sectors. Indeed the mass of the singlet dominated Higgs scalar 
particle $m_{h_2}$ is always close to the mass of $Z'$ boson 
$M_{Z'}\simeq g^{'}_1 \tilde{Q}_S s\sim g_1 s$ that has to be heavier than 
$500-600\,\mbox{GeV}$ \cite{5}. The masses of the charged, CP--odd and one CP--even 
Higgs states are governed by $m_A$. The mass of the SM--like Higgs boson given by 
$Eq.(\ref{6})$ is set by $M_Z$. The last term in Eq.(6) must not be allowed to dominate 
since it is negative. This constrains $x$ around unity for $\lambda>g'_1$.
As a consequence $m_A$ is confined in the vicinity of $\frac{\lambda s}{\sqrt{2}}\,\tan\beta$ and 
is much larger than the masses of the $Z'$ and $Z$ bosons. At so large values of $m_A$ the 
masses of the heaviest CP--even, CP--odd and charged states are almost degenerate around $m_A$.

The qualitative pattern of the Higgs spectrum obtained for $\lambda>g'_1$ in the leading 
one--loop approximation is shown in Fig.~1. The numerical analysis reveals that the heaviest
CP--even, CP--odd and charged Higgs states lie beyond the $\mbox{TeV}$ range in the 
considered case. The second lightest CP--even Higgs boson is predominantly a singlet field 
so that it will be quite difficult to observe this particle at future colliders. 
With decreasing $\lambda$ the allowed range of $m_A$ enlarges so that charged, CP--odd and
second lightest CP--even Higgs states may have masses in the $200-300\,\mbox{GeV}$ range
when $\lambda<g'_1$. But for $m_A< 500\,\mbox{GeV}$ and $\lambda<g'_1$ we get a MSSM--type
Higgs spectrum with the lightest SM--like Higgs boson below $130-135\,\mbox{GeV}$ and 
with the heaviest scalar above $600\,\mbox{GeV}$ being singlet dominated and irrelevant. 
The non--observation of Higgs particle at LEP rules out most parts of the E$_6$SSM parameter 
space in this case.
\begin{figure}
\hspace{-0cm}{\large $m_{h_i}$}\\[2mm]
\epsfxsize=280pt
\parbox{\epsfxsize}{\epsfbox{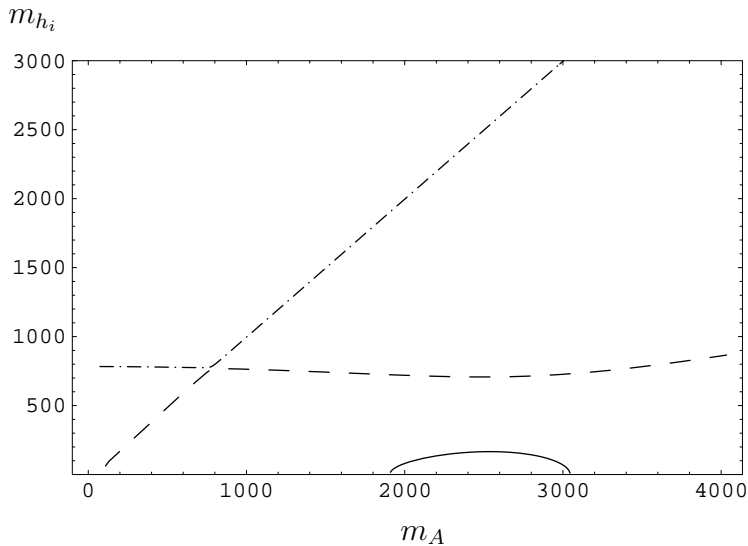}}\\
\begin{center}
{\large $m_A$~~~~~~~~}
\end{center}
\caption{The one--loop masses of the CP--even Higgs bosons versus $m_A$ for
$\lambda(M_t)=0.794$, $\tan\beta=2$, $M_{Z'}=M_S=700\,\mbox{GeV}$ and $X_t=\sqrt{6}M_S$.  
Solid, dashed and dashed--dotted lines correspond to the masses of the lightest, 
second lightest and heaviest Higgs scalars.}
\end{figure}

From Fig.~1 and Eq.~(\ref{6}) it becomes clear that at some value of $m_A$ (or $x$)
the lightest CP--even Higgs boson mass $m_{h_1}$ attains its maximum value. At 
tree--level the upper bound on $m_{h_1}$ is given by the sum of the first three
terms in Eq.~(\ref{6}). The inclusion of loop corrections increase the bound on the 
lightest CP-even Higgs boson mass in SUSY models substantially. In Fig.~2 we plot
the two-loop upper bounds on the mass of the lightest Higgs particle in the MSSM, 
NMSSM and E$_6$SSM as a function of $\tan\beta$. At moderate values of $\tan\beta$ 
($\tan\beta=1.6-3.5$) the upper limit on the lightest Higgs boson mass in the E$_6$SSM 
is considerably higher than in the MSSM and NMSSM. It reaches the maximum value 
$\sim 150-155\,\mbox{GeV}$ at $\tan\beta=1.5-2$. In the considered part of the 
parameter space the theoretical restriction on the mass of the lightest CP-even 
Higgs boson in the NMSSM exceeds the corresponding limit in the MSSM because
of the extra contribution to $m^2_{h_1}$ induced by the additional $F$-term in the 
Higgs scalar potential of the NMSSM. The size of this contribution, which is 
described by the first term in Eq.~(\ref{6}) is determined by the Yukawa coupling
$\lambda$.  The upper limit on the coupling $\lambda$ caused by the validity of 
perturbation theory in the NMSSM is more stringent than in the E$_6$SSM due to the 
presence of exotic $5+\overline{5}$-plets of matter in the particle spectrum of 
the E$_6$SSM. This is the reason why the upper limit of $m_{h_1}$ in the NMSSM is 
considerably less than in the E$_6$SSM at moderate values of $\tan\beta$.
\begin{figure}
\hspace{-0cm}{\large $m_{h_1}$}\\[2mm]
\epsfxsize=280pt
\parbox{\epsfxsize}{\epsfbox{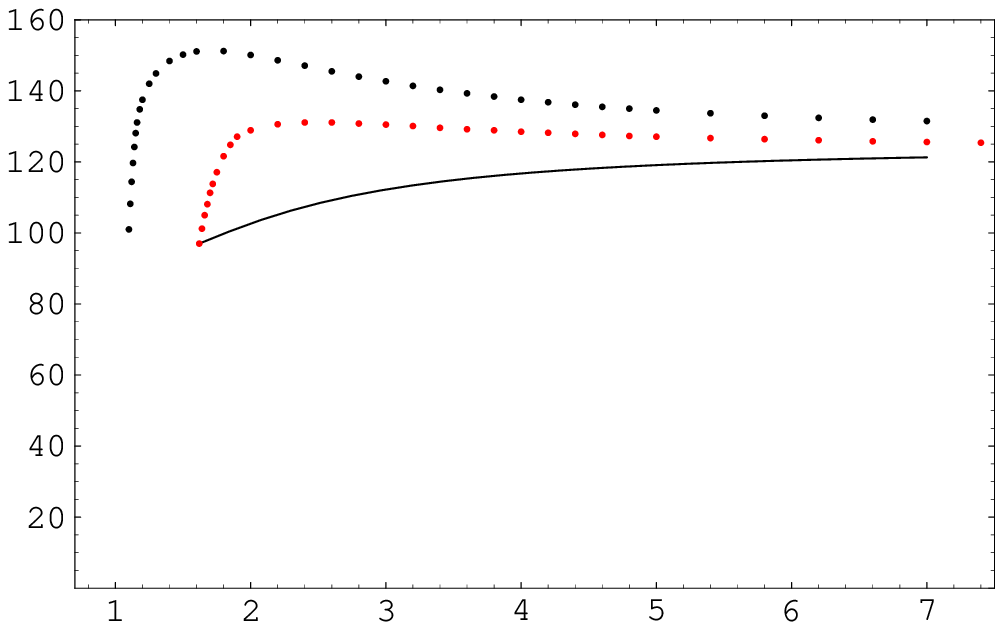}}\\
\begin{center}
{\large $\tan\beta$~~~~~~~~}
\end{center}
\caption{The dependence of the two-loop upper bound on the lightest
Higgs boson mass on $\tan\beta$ for $m_t(m_t)=165\,\mbox{GeV}$, $m_Q^2=m_U^2=M_S^2$,
$X_t=\sqrt{6} M_S$ and $M_S=700\,\mbox{GeV}$. The solid, lower and upper dotted lines
represent the limit on $m_{h_1}$ in the MSSM, NMSSM and E$_6$SSM.}
\end{figure}

At large $\tan\beta> 10$ the contribution of the $F$-term of the SM--type 
singlet field to $m_{h_1}^2$ vanishes. Therefore with increasing $\tan\beta$ the 
upper bound on the lightest Higgs mass in the NMSSM approaches the 
corresponding limit in the MSSM. In the E$_6$SSM the theoretical restriction on 
$m_{h_1}$ also diminishes when $\tan\beta$ rises but it is still 
$4-5\,\mbox{GeV}$ larger than the one in the MSSM because of the $U(1)_{N}$ 
$D$-term contribution to $m^2_{h_1}$ (the third term in Eq.~(\ref{6})).

The discovery at future colliders of superpartners of observed quarks and leptons as 
well as a relatively heavy SM--like Higgs boson with mass $140-155\,\mbox{GeV}$, that 
corresponds to $\lambda>g'_1$ in the E$_6$SSM, will permit to distinguish the E$_6$SSM from 
the simplest supersymmetric (like MSSM and NMSSM) and other extensions of the SM.
Another possible manifestations of the exceptional SUSY model at the LHC are
related with the enhanced production of $l^{+}l^{-}$, $t\bar{t}$ and/or $b\bar{b}$ pairs 
coming from either a $Z'$ boson or exotic particle decays.

\vspace*{0.2cm}\noindent
{\bf Acknowledgements:}~ The authors would like to to acknowledge support from the 
PPARC grant PPA/G/S/2003/00096, the NATO grant PST.CLG.980066 and the EU network 
MRTN 2004-503369.

\renewcommand{\baselinestretch}{1.00}

\end{document}